\documentclass[prd,nofootinbib,amsfonts,notitlepage]{revtex4-2}
\usepackage[utf8]{inputenc}
\usepackage[T1]{fontenc}
\usepackage[colorlinks, linkcolor = red]{hyperref}
\usepackage{graphicx,bm,amsmath,color}
\usepackage{graphicx,bm,amsmath,color,scalerel}
\usepackage{bbold}
\usepackage{wasysym}
\usepackage{verbatim}
\usepackage{amssymb}
\usepackage{epstopdf}
\usepackage{xmpmulti}
\usepackage{centernot}
\usepackage{multirow}
\usepackage{tikz}
\usepackage{graphicx}
\usepackage{float}
\usepackage{mathtools}
\usepackage{comment}
\makeatletter

\newcommand{\be}{\begin{equation}}
	\newcommand{\ee}{\end{equation}}
\newcommand{\beq}{\begin{eqnarray}}
	\newcommand{\eeq}{\end{eqnarray}}
\newcommand{\ba}{\begin{align}}
	\newcommand{\ea}{\end{align}}

\begin{document}
    \title{Deformed relativistic kinematics on curved spacetime - a geometric approach}

	\author{Christian Pfeifer}
	\email{christian.pfeifer@zarm.uni-bremen.de}
	\affiliation{ZARM, University of Bremen, 28359 Bremen, Germany}
	
	\author{José Javier Relancio}
	\email{relancio@unizar.es}
	\affiliation{Dipartimento di Fisica ``Ettore Pancini'', Università di Napoli Federico II, 80138 Napoli, Italy;\\
	INFN, Sezione di Napoli, 80126 Napoli, Italy;\\
	Centro de Astropartículas y Física de Altas Energías (CAPA), Universidad de Zaragoza, Zaragoza 50009, Spain}

\begin{abstract}
	Deformed relativistic kinematics have been considered as a way to capture residual effects of quantum gravity. It has been shown that they can be understood geometrically in terms of a curved momentum space on a flat spacetime. In this article we present a systematic analysis under which conditions and how deformed relativistic kinematics, encoded in a momentum space metric on flat spacetime, can be lifted to curved spacetimes in terms of a self-consistent cotangent bundle geometry, which leads to purely geometric, geodesic motion of freely falling point particles. We comment how this construction is connected to, and offers a new perspective on, non-commutative spacetimes. From geometric consistency conditions we find that momentum space metrics can be consistently lifted to curved spacetimes if they either lead to a dispersion relation which is homogeneous in the momenta, or, if they satisfy a specific symmetry constraint. The latter is relevant for the momentum space metrics encoding the most studied deformed relativistic kinematics. For these, the constraint can only be satisfied in a momentum space basis in which the momentum space metric is invariant under linear local Lorentz transformations. We discuss how this result can be interpreted and the consequences of relaxing some conditions and principles of the construction from which we started.
\end{abstract}

\maketitle


\section{Introduction}
Due to a missing self-consistent theory of quantum gravity (QG), and the unsolved tensions between quantizing general relativity (GR) and the standard approaches of how to quantize physical field theories, models which try to capture expected features of the quantum nature of gravity have been brought forward. Among them are modified relativistic kinematics (MRKs), which describe the interaction of particles with QG effectively below the Planck scale~\cite{Liberati:2013xla,AmelinoCamelia:2008qg,AmelinoCamelia:2011pe,AmelinoCamelia:2011bm}. Versions of MRKs have already been derived from fundamental approaches to QG like Loop QG~\cite{Assanioussi:2014xmz,Brahma:2018rrg} and string theory~\cite{Ellis:1999uh}.

The main idea behind MRKs is that high-energetic point particles are able to probe smaller distances than low-energetic particles. Assuming that the scale of quantum gravity is a high-energy (small-distance) scale $\Lambda$, often identified with the Planck scale, higher-energetic probe particles should reveal more information about the physics at the QG scale than low-energetic ones. Since classical gravity is described by a curved spacetime, this idea can effectively be modeled by a four-momentum dependent spacetime geometry. In general, this structure is not necessarily local Lorentz invariant; a statement which does not say anything about if, or if not, the yet to be found fundamental theory of QG is local Lorentz invariant. Similar approaches are known from the study of particles and fields in media. Phenomenologically, their behavior can be described by a non Lorentz invariant background geometry, even though their interactions with the fundamental constituents of the medium are governed by the local Lorentz invariant standard model of particle physics.

Among the MRKs one distinguishes between two scenarios: Lorentz invariance violation (LIV) and deformed relativistic kinematics (DRKs). In the LIV case, the information about the MRKs are encoded in a, compared to GR, modified dispersion relation, which is satisfied by particles propagating through spacetime. However, observers are related to each other by local Lorentz transformations, and the observer momenta obey the general relativistic dispersion relation. Since the dispersion relation encodes the coupling between the physical systems and the space-time geometry, this distinction between particles and observers violates the weak equivalence principle, the fact that gravity couples in the same way, universally, to all physical systems. In contrast, in the DRKs case, all physical systems (point particles and observers) satisfy the same dispersion relation, and, most importantly, also a compatible deformed addition of momenta is implemented~\cite{AmelinoCamelia:2008qg}. Thus, for DRKs, the information about the deviations from local Lorentz invariance are not only encoded in a possibly deformed dispersion relations, but also in the deformed observer transformations and in the composition of momenta. This construction ensures a relativity principle compatible with a four momentum dependent geometry of spacetime.

DRKs are often constructed by deforming the Poincaré symmetry algebra transformations of special relativity acting on Minkowski spacetime. A most prominent example of such a deformation is the $\kappa$-Poincaré Hopf algebra~\cite{Lukierski:1992dt}, which is interpreted as a symmetry principle that implements two observer independent scales: a velocity, which corresponds to the special relativistic speed of light ($c$) and, in addition, a characteristic high-energy scale ($\Lambda$), usually identified with the Planck energy. In this sense, physics obeying this symmetry has also been named doubly special relativity (DSR)~\cite{AmelinoCamelia:2010pd}. Instead of the algebraic approach to DRKs, and in particular DSR, a geometric interpretation has been suggested, namely that the deformed symmetries are encoded in a non-trivial geometry of momentum space~\cite{KowalskiGlikman:2002ft,AmelinoCamelia:2011pe,AmelinoCamelia:2011bm,Barcaroli:2015eqe,Carmona:2019fwf}. This works well on flat spacetimes, but a self-consistent generalization to curved spacetimes is still missing.

The step from flat to curved spacetime is highly important, since the most promising observables to detect consequences from DRKs come from cosmic messengers. One well discussed effect is a time delay in time of arrival measurements: high-energetic particles emitted simultaneously from an astrophysical source at high redshift accumulate DRK effects due to their long travel time \cite{AmelinoCamelia:1997gz,Amelino-Camelia1998,Ellis:1999uh,Jacob:2008bw,Martinez:2008ki,AmelinoCamelia:2011cv,Freidel:2011,Rosati:2015pga,Barcaroli:2016yrl,Carmona:2017oit,Pfeifer:2018pty}. This leads to a possibly measurable deviation from the expected time of arrival predicted by GR. But, in order to describe this effect self-consistently, a consistent complete formulation of DRKs on a curved spacetime is necessary.

A further important aspect in the formulation of DRKs on curved spacetime is the definition of the trajectories of particles. Usually, these are defined from the Hamilton equations, considering a deformed Hamilton function which encodes the modified dispersion relation. However, as explained previously, the main ingredient of DRKs is the deformed composition law for the momenta constructed from a momentum space metric. Whilst one way to construct momentum dependent metrics on curved spacetime is to derive them from the Hamilton function~\cite{Barcaroli:2016yrl,Barcaroli:2015xda,Barcaroli:2017gvg}, a formalism known as Hamilton geometry~\cite{miron2001geometry}, the connection with a deformed addition of momenta is not clearly worked out in this context  yet. An alternative approach is to start from a momentum space metric on flat spacetime, instead of from a dispersion relation. Then, the deformed addition of momentum, as well as the deformed observer transformations, can be understood as the isometries of a maximally symmetric momentum space~\cite{Carmona:2019fwf}.  For such a momentum space metric, the Hamiltonian is defined as the square of the metric distance in momentum space. The mathematical framework which covers this approach is the geometry of generalized Hamilton spaces~\cite{miron2001geometry}, and first steps to implement DRKs on curved spacetime in this framework have been made in~\cite{Relancio:2020zok,Relancio:2020rys}. In particular several consistency conditions have been identified, but not generally studied yet. Here we seek to investigate systematically which kind of curved spacetime geometries with curved momentum spaces, collectively described as a cotangent bundle geometry, are compatible with these consistency conditions. We find that, among the geometries which emerge from lifting DRKs from flat to curved spacetimes, only specific classes satisfies all consistency conditions: those which have a dispersion relation that is a homogeneous function of the momenta, or, those which are linearly local Lorentz invariant. One can construct more general self-consistent curved spacetimes with curved momentum spaces geometry, but they then either violate one of the conditions we started from, or go beyond the DRKs usually discussed in the literature.

Another point that deserves discussion is the connection between this kind of geometrical structure and the quantum gravity framework. As discussed in~\cite{AmelinoCamelia:2008qg}, DSR theories have a long standing history in the literature since their emergence from quantum deformations of the Poincar\'e group \cite{Lukierski:1991pn,Lukierski:1993df,Lukierski:1992dt,Lukierski:2002df}, whose physical interpretation is nicely summarized in~\cite{Amelino-Camelia2002b}. Not long after that it was recognized in~\cite{Freidel:2003sp} that such a framework could be rigorously derived from a top down approach in $2+1$ quantum gravity. The year after it was suggested that DSR could be the outcome of an energy (rainbow) spacetime~\cite{Magueijo:2002xx}, showing therefore a clear connection between a momentum dependent spacetime and quantum gravity. A quantum spacetime it is often described by a non-commutativity of spacetime. Following the same line of thought  of~\cite{Carmona:2021gbg,Relancio:2021ahm} (see also~\cite{Wagner:2021bqz}), we show how to connect our geometrical setup with a space-time non-commutativity.  Explicitly, starting by a momentum dependent geometry and we can identify the space-time coordinates with the generators of translations in momentum space, leading to a non-commutative phase space.

This differs from the construction of~\cite{Ballesteros:2017pdw,Ballesteros:2017kxj,Ballesteros:2019hbw}, where it is extended  the usual construction of $\kappa$-Poincaré Hopf algebras when a cosmological constant in spacetime is considered, and the approach followed in~\cite{Beggs:2013pxa,Doplicher:1994zv}, where a manifold with non-commutative coordinates was considered from the very beginning.  The construction proposed here has the advantage that it is easily generalizable for any curved space-time geometry.

The structure of this article is as follows. We start by recalling the geometric understanding of deformed relativistic kinematics on flat spacetime and conjecture how to extend the deformed relativistic kinematics to curved spacetime in Sec.~\ref{sec:DRKsFlatToCurved}, including a discussion on the relation of this construction to non-commutative spacetimes in Sec.~\ref{sec:noncom}. Afterwards, we introduce the geometry of the cotangent bundle and state two physical principles, which we want to be satisfied by the geometry in Sec.~\ref{sec:GeomT*M}. Using the framework of the geometry of the cotangent bundle we cast these principles in precise mathematical constraints in Sec.~\ref{ssec:geomMet}. Then, in Sec.~\ref{ssec:nT*Mgeom} we study the consequences of the constraints on a perturbative model which is polynomial in the momenta, as they are often employed in the study of DRKs, and find that the perturbation of a space-time metric to a momentum space metric must satisfy an index symmetry constraint in order to yield a consistent geometry. Finally we evaluate this constraint for the momentum space metrics of the most popular DRKs in Sec.~\ref{sec:geomex}, where we find that it is not satisfied by every momentum basis, i.e., every choice of momentum coordinates, but only for those maximally symmetric metrics which have linear Lorentz transformations as isometries. Finally, we discuss the main outcomes of the work in Sec.~\ref{sec:conclusions}, including the interpretation of our findings and possible generalisations.

\section{Deformed relativistic kinematics: from flat to curved spacetimes}\label{sec:DRKsFlatToCurved}
Deformed relativistic kinematics (DRKs) can be understood in terms of non-trivial curved momentum space geometry, defined by a momentum space metric. One of the aims of this article is to develop the geometric notions to lift this model consistently to curved spacetime. To do so, we recall the flat spacetime construction and present a simple way for lifting this setup to curved spacetime. In Secs. \ref{sec:GeomT*M} and \ref{sec:geomex} we will study under which conditions the construction yields a mathematically self-consistent curved phase-space geometry, i.e., a curved spacetime with curved momentum spaces.

\subsection{The flat spacetime construction}\label{sec:DRKsFlat}
Originally, curved momentum spaces have been introduced by Born~\cite{Born:1938} to unify quantum theory and relativity. In the geometric approach to DRKs, momentum spaces are equipped with a maximally symmetric metric $\zeta = \zeta^{\mu\nu}(k) dk_\mu dk_\nu$. The maximal symmetry requirement implies the existence of $10$ isometries, which consist of $4$ translations $\mathfrak{T}$ and $6$ boosts and rotations $\mathfrak{J}$ (we are considering metrics of Lorentzian signature and $4$ dimensions). Along this paper, we will restrict ourselves to this particular kind of momentum spaces, since we will identify translations and Lorentz generators with the deformed law of addition of momenta and the rotations and boost respectively~\cite{Carmona:2019fwf}. For momentum spaces that are not maximally symmetric this construction cannot be carried out, failing then to identify geometrically the relativistic deformed kinematics.

The dispersion relation, which physical momenta have to satisfy, is obtained as the square of the minimal geometric distance of a momentum $k$ from the origin of momentum space, measured by the momentum space length measure induced by the metric~\cite{AmelinoCamelia:2011bm}. As discussed before, translations induce a deformed law of addition of momenta, and the rotations and boosts represent the local observer transformations~\cite{Carmona:2019fwf}
\begin{align}\label{eq:DRKflat}
	(p\oplus q)_\mu \,=\, \mathfrak{T}(p,q)_\mu \,,\quad p^\prime_\mu \,=\, \mathfrak{J}(p,\Omega)_\mu\,,
\end{align}
where $\mathfrak{T}(p,q)$ is the translation of $p$ by another momentum $q$ (used as parameter of the translation) to the momentum $(p\oplus q)_\mu$, and $\mathfrak{J}(p,\Omega)$ are the rotations and boosts, parametrized by the  matrix $\Omega_{\mu\nu} = -\Omega_{\nu\mu}$.

The isometries are generated by the momentum space vector fields 
\begin{equation}\label{eq:generators_flat}
	\mathcal{T}^\mu\,= \, T^\mu{}_\lambda (k)\frac{\partial}{\partial k_\lambda}\,,\qquad \mathcal{J}^{\mu\nu}\,=\,  J^{\mu\nu}{}_\lambda (k)\frac{\partial}{\partial k_\lambda}\,,
\end{equation}  
which define the deformed symmetry algebra, and are Killing vector fields of the momentum space metric. Their action on momenta is
\begin{align}
	\tilde{p}_\mu  \,=\,p_\mu+ q_\nu T^\nu{}_\mu(p)+\Omega_{\rho\sigma} J^{\rho\sigma}{}_\mu(p) \,.
	\label{eq:prime_p}
\end{align} 

An explicit representation of a maximally symmetric momentum space metric, in a global Cartesian coordinate on flat spacetime, is given by
\begin{align}\label{eq:gmaxflat}
	\zeta^{\mu\nu}(k) = \eta^{\mu\nu} + \frac{K}{(1-K \eta^{\rho\sigma}k_\sigma k_\rho)}\eta^{\mu\lambda}k_\lambda \eta^{\nu\iota}k_\iota\,,
\end{align}
where $\eta$ is the Minkowski metric, and the curvature parameter $K$ is identified with the QG scale by $K=\pm 1/\Lambda^2$. The generators of translations, boosts, and rotations, are
\begin{align}
	\label{eq:gensymflat}
	\mathcal{T}^\lambda \,=\,\sqrt{1 - K \eta^{\mu\nu}k_\mu k_\nu}\ \frac{\partial}{\partial k_\lambda}\,, \quad  \mathcal{J}^{\mu \nu} \,=\, k_\rho(\delta^\mu_\lambda \eta^{\nu\rho}-\delta^\nu_\lambda \eta^{\mu\rho}) \frac{\partial}{\partial k_\lambda}\,.
\end{align}
They form a de-Sitter or anti-de-Sitter algebra for $K$ positive or negative, respectively. For $K=0$, one recovers the Poincar\'e algebra. 

The representation of the symmetry generating vector fields chosen here defines the so-called Snyder algebra DRKs~\cite{Carmona:2019fwf}. Different DRKs are obtained by defining new translation generators as
\begin{align}\label{eq:transredef}
	\tilde{\mathcal{T}}^\mu \,= \,\mathcal{T}^\mu +c^\mu{}_{\nu \rho}\, \mathcal{J}^{\nu\rho}\,,
\end{align}
where it depends on the choice of the coefficients $c^\mu{}_{\nu \rho}$ which kind of DRKs one constructs. These redefinitions of the translation generators are equivalent to a redefinition of the momentum composition law associated to them.

For example, the $\kappa$-Poincar\'e algebra is obtained from a de-Sitter metric when the translation generators form a closed subalgebra, which means choosing $c^\mu{}_{\nu \rho} = \delta^\mu_\nu n_\rho/\Lambda$~\cite{Carmona:2019fwf}\footnote{The $\kappa$-Minkowski algebra can be only obtained from the generators of a de Sitter momentum space, being impossible to find it for anti-de Sitter~\cite{Carmona:2019fwf}.}, where $\Lambda$ is usualy interpreted as a high energy scale and $n_\rho$ are the components of a timelike normalized covector on Minkowski spacetime, $\eta^{\mu\nu}n_\mu n_\nu=-1$, that can be chosen to be $n_\mu = (1,0,0,0)$. Explicitly the new generators of translations in momentum space are easily calculated to be
\begin{align}\label{eq:transkappa}
	\tilde{\mathcal{T}}^\mu = \mathcal{T}^\mu + \frac{n_\rho}{\Lambda} \mathcal{J}^{\mu\rho} =  \mathcal{T}^\mu + \frac{1}{\Lambda}\mathcal{J}^{\mu 0}\,.  
\end{align}

\subsection{Non-commutative spacetime from isometry generators}\label{sec:noncom}

Before we lift the deformed kinematics to curved spacetime, we like to point out their relation to non-commutative spacetimes. 

As demonstrated in the previous section, and shown in~\cite{Carmona:2019fwf}, the kinematics of $\kappa$-Poincaré~\cite{Majid1994}, Snyder~\cite{Battisti:2010sr}, and the so-called hybrid models~\cite{Meljanac:2009ej}, can be obtained from the geometrical ingredients of a maximally symmetric momentum space.

It is straightforward to calculate the commutator relations between the generators of the isometries~\eqref{eq:gensymflat}. They form the algebra~\cite{Battisti:2010sr,Carmona:2019fwf} 
\be
\begin{split}
	[\mathcal{T}^\alpha, \mathcal{T}^\beta] \,=\,& K \mathcal{J}^{\alpha\beta}\,, \quad [\mathcal{T}^\alpha, \mathcal{J}^{\beta\gamma}]\,=\, \eta^{\alpha\beta} \mathcal{T}^\gamma - \eta^{\alpha\gamma} \mathcal{T}^\beta\,, \\ [\mathcal{J}^{\alpha\beta},\mathcal{J}^{\gamma\delta}]\,=\,& \eta^{\beta\gamma}\mathcal{J}^{\alpha\delta} - \eta^{\alpha\gamma}\mathcal{J}^{\beta\delta} - \eta^{\beta\delta}\mathcal{J}^{\alpha\gamma} + \eta^{\alpha\delta}\mathcal{J}^{\beta\gamma}\,.
\end{split}
\label{cov_generators}
\ee

Now we can make the identification of the non-commutative space-time coordinates with the generators of translations, viz.
\be
\bar{x}^\alpha \,=\,\mathcal{T}^\alpha\,,
\label{eq:flat_x}
\ee
from which we can read the following commutators
\be
\begin{split}
	[\bar{x}^\alpha ,\bar{x}^\beta ]\,=\,&- K  \mathcal{J}^{\alpha\beta}\,,\,\,
	[\bar{x}^\alpha, k_\beta]\,=\,\delta^\alpha_\beta \sqrt{1 -K  \eta^{\mu\nu}k_\mu k_\nu} \,,\\ 
	[\bar{x}^\alpha, \mathcal{J}^{\beta\gamma}]\,=\,& \eta^{\alpha\beta} \bar{x}^\gamma - \eta^{\alpha\gamma} \bar{x}^\beta\,.
\end{split}
\label{eq:flat_ps}
\ee
This phase space corresponds to the Snyder phase-space non-commutativity~\cite{Battisti:2010sr} and the composition law (defined as the finite translations)~\cite{Carmona:2019fwf}
\be
(p\oplus q)_\mu \,= \, p_\mu \left(\sqrt{1- K q_\mu q_\nu \eta^{\mu\nu}}-
\frac{K p_\mu q_\nu \eta^{\mu\nu}}{ \left(1+\sqrt{1-K p_\mu p_\nu \eta^{\mu\nu}}\right)}  \right) + q_\mu \,,
\label{DCLSnyder-1}
\ee
of the Maggiore representation. Different non-commutative flat spacetimes can be obtained by a redefinition of the translation generators, as explained below~\eqref{eq:transredef}. For  example, for the change to the $\kappa$-Minkowski translations \eqref{eq:transkappa}  one easily finds
\be
\begin{split}
	[ \tilde{\mathcal{T}}^\alpha,  \tilde{\mathcal{T}}^\beta] \,=\,&\frac{n_\gamma}{\Lambda}\left( \tilde{\mathcal{T}}^\alpha \eta^{\beta\gamma} - \tilde{\mathcal{T}}^\beta \eta^{\alpha\gamma} \right) \,,\\ [ \tilde{\mathcal{T}}^\alpha, \mathcal{J}^{\beta\gamma}]\,=\,& \eta^{\alpha\beta}  \tilde{\mathcal{T}}^\gamma - \eta^{\alpha\gamma}  \tilde{\mathcal{T}}^\beta+\frac{n_\delta}{\Lambda}\left(\eta^{\delta\beta} \mathcal{J}^{\alpha\gamma}-\eta^{\delta\gamma} \mathcal{J}^{\alpha\beta}\right)\,.
\end{split}
\label{cov_algebra_kappa}
\ee

\subsection{Lifting deformed relativistic kinematics to curved spacetime}\label{ssec:DRKsCST}
To study predictions from deformed relativistic kinematics on curved spacetimes, such as the existence or absence of energy dependent time delays in time of arrival measurement of  high-energetic photons from cosmological distances \cite{Jacob:2008bw,Rosati:2015pga,Barcaroli:2016yrl,Amelino-Camelia:2016ohi,Pfeifer:2018pty}, photon orbits \cite{Barcaroli:2017gvg}, and other gravitational lensing observations \cite{Glicenstein:2019rzj}, it is necessary to extend the previous discussion to a curved spacetime scenario. For these observables, curved space-time effects have an essential influence and cannot be neglected.

Moreover, from the geometrical approach, the step from flat to curved spacetimes is naturally the next one, since the gravitational interaction can be described by a curved spacetime geometry. We have seen that the passage from SR to DSR is depicted by a curved momentum space, so we should be able to combine both curvatures in order to obtain a deformation of GR. In this way, we would be able to describe deformed kinematics on a curved spacetime from a geometrical point of view~\cite{Cianfrani:2014fia,Amelino-Camelia:2014rga,Barcaroli:2015xda,Lobo:2016xzq,Letizia:2016lew,Barcaroli:2017gvg,Relancio:2020zok}.

For that aim, we implement the DRKs on curved spacetimes by localizing the momentum space geometry to each point in spacetime. \vspace{10pt}

\textbf{Conjecture:} \emph{Given a spacetime $M$ equipped with a Lorentzian metric $a$ with local coordinate components $a_{\mu\nu}(x)$, then its QG scale deformed momentum dependent geometry is determined by the position dependent momentum space metric with local coordinate components
	\begin{align}\label{eq:gmaxcurved}
		g^{\mu\nu}(x,k) = a^{\mu\nu}(x) + \frac{K}{(1-K a^{\rho\sigma}(x)k_\sigma k_\rho)}a^{\mu\lambda}(x)k_\lambda a^{\nu\iota}(x)k_\iota\,.
	\end{align}
}

In fact, we can decompose the spacetime metric as a function of the components of a tetrad $e^\mu{}_\nu = e^\mu{}_\nu(x)$ of the metric, i.e.,\ $a^{\mu\nu}(x) = e^\mu{}_\alpha \eta^{\alpha\beta} e^\nu{}_\beta$, so that the previous equation can be written as 
\begin{equation}
	g^{\mu \nu}(x,k)\,=\,e^{\mu}{}_\alpha \eta^{\alpha \beta}e^{\nu}{}_\beta +\frac{K}{1- K e^{\lambda}{}_\iota \eta^{\iota\epsilon} e^{\sigma}{}_\epsilon k_\lambda k_\sigma} e^{\mu}{}_\alpha \eta^{\alpha \beta}e^{\lambda}{}_\beta k_\lambda e^{\nu}{}_\gamma \eta^{\gamma \delta} e^{\sigma}{}_\delta k_\sigma \,.
\end{equation}
This shows that the curved spacetime and momentum space metric \eqref{eq:gmaxcurved} assumes its flat spacetime form \eqref{eq:gmaxflat} in the frames of the metric $a$. Defining $\bar{k}_\alpha= e^{\mu}{}_\alpha k_\mu$, we can rewrite the previous expression easily as 
\begin{equation}
	g^{\mu \nu}(x,k)\,=\,e^{\mu}{}_\alpha \zeta^{\alpha \beta}(\bar{k})e^{\nu}{}_\beta \,.
	\label{eq:definition_metric_cotangent}
\end{equation}
This is in agreement with what we found in~\cite{Relancio:2020zok}.

The generators of isometries on curved spacetime become space-time dependent and can be obtained from their flat spacetime counterparts~\eqref{eq:gensymflat}, either by a frame transformation $k_\mu \to k_\nu e^\nu{}_\mu$ of the momenta, or, equivalently, by the replacement of the flat by the curved spacetime metric $\eta^{\mu\nu}\to a^{\mu\nu}(x)$. The DRKs are then defined at each spacetime point, analogously as they are defined on flat spacetime~\eqref{eq:DRKflat}.

General coordinate transformations on momentum space yield the momentum space geometry which represents DRKs on curved spacetime in different bases~\cite{Relancio:2020zok,Relancio:2020rys}. In the new basis the resulting momentum space metric may not only depend on a spacetime metric, but in general on further space-time tensors, such as additional vector fields. In the literature, often one additional vector field is employed. We will discuss this possibility again in Sec. \ref{sec:geomex}. The phenomenological consequences may be different, for different bases or different choices of translations generators, as it is already the case on flat spacetime.

Alternatively one obtains different DRKs by redefining the generators of translation, analogously as on flat spacetime,
\begin{align}\label{eq:transredef2}
	\tilde{\mathcal{T}}^\mu \,=\, \mathcal{T}^\mu +c^\mu{}_{\nu \rho}(x)\, \mathcal{J}^{\nu\rho}\,,
\end{align}
with the difference that on curved spacetimes the coefficients, which define the translation generators, may depend on the space-time point $c^\mu{}_{\nu \rho}(x)$.  Hence, there may be space-time regions where they differ, and even vanish. This gives the possibility to construct spacetimes with regions where different DRKs are relevant.

To transform the local Snyder algebra~\eqref{eq:gensymflat}, for example to a local $\kappa$-Poincar\'e algebra in the classical basis on curved spacetimes, one needs to choose $c^\mu{}_{\nu \rho}(x) = \delta^\mu_\nu Z_\rho(x)/\Lambda$, where $Z_\alpha(x)=n_\nu {e^\nu}_\alpha(x)$ and $a^{\rho\sigma}(x)Z_\rho(x)Z_\sigma(x) = -1$. Considering a vector field with compact support, which can be constructed by multiplication of the components $Z_\alpha(x)$ of any unit timelike $1$-form by a function of compact support, would directly lead to a spacetime with regions of different DRKs.

In this case then, in the regions where $Z_\sigma(x)\neq0$, the translation generators become
\begin{align}\label{eq:transkappacurved}
	\tilde{\mathcal{T}}^\mu = \mathcal{T}^\mu + \frac{Z_\rho(x)}{\Lambda} \mathcal{J}^{\mu\rho}\,.
\end{align}
We introduce this new possibility of equipping spacetimes with different DRKs in different regions here for the first time and will investigate it in detail in future work.

Before we continue to study the geometric consistency and consequences of this just outlined geometric construction for deformed relativistic kinematics on curved spacetimes in the next section \ref{sec:GeomT*M}, we like to comment on the interpretation of the deformed relativistic kinematics as non-commutativity of spacetime in the curved case. 

\subsection{Local non-commutative spacetime from isometry generators on curved spacetimes}\label{sec:noncomCST}
On curved spacetimes the generators of isometries in momentum space, at each point of spacetime become, 
\begin{align}\label{eq:generators_desitter_curved}
	\mathcal{T}^\lambda(x) \,=\, \sqrt{1 - K k^2 } \ \frac{\partial}{\partial k_\lambda}\,,   \qquad 
	\mathcal{J}^{\mu \nu}(x) \,=\, k_\rho(\delta^\nu_\lambda a^{\mu\rho}-\delta^\mu_\lambda a^{\nu\rho}) \frac{\partial}{\partial k_\lambda}\,,
\end{align}
where  $k^2=k_\mu k_\nu a^{\mu \nu}(x)$, and satisfy
\be
\begin{split}
	[\mathcal{T}^\alpha, \mathcal{T}^\beta] \,=\,& K \mathcal{J}^{\alpha\beta}\,, \quad [\mathcal{T}^\alpha, \mathcal{J}^{\beta\gamma}]\,=\, \eta^{\alpha\beta} \mathcal{T}^\gamma - \eta^{\alpha\gamma} \mathcal{T}^\beta\,,\\
	[\mathcal{J}^{\alpha\beta},\mathcal{J}^{\gamma\delta}]\,=\,& \eta^{\beta\gamma}\mathcal{J}^{\alpha\delta} - \eta^{\alpha\gamma}\mathcal{J}^{\beta\delta} - \eta^{\beta\delta}\mathcal{J}^{\alpha\gamma} + \eta^{\alpha\delta}\mathcal{J}^{\beta\gamma}\,.
\end{split}
\label{cov_generators_curved}
\ee

On a curved spacetime, the identification of the generators of translations on-momentum space with the coordinates of a curved non-commutative spacetime is not as straightforward as on flat spacetime.

One possibility is to identify again the non-commutative spacetime with the generators of translations, $\bar x^\alpha(x,k) = \mathcal{T}^\alpha(x,k)$, and again we find (where we suppressed the $x$-dependence on $\mathcal{J}^{\mu \nu}=\mathcal{J}^{\mu \nu}(x)$),
\be
[\bar{x}^\alpha ,\bar{x}^\beta ]\,=\,- K  \mathcal{J}^{\alpha\beta}\,,\quad 
[\bar{x}^\alpha, k_\beta]\,=\,\delta^\alpha_\beta \sqrt{1-K   k^2} \,,\quad  
[\bar{x}^\alpha, \mathcal{J}^{\beta\gamma}]\,=\, a^{\alpha\beta} \bar{x}^\gamma - a^{\alpha\gamma} \bar{x}^\beta\,,
\label{eq:curved_ps}
\ee
and in this case the composition law $\bar{\oplus}$ for a generic curved spacetime reads 
\be
(p\oplus q)_\mu \,= \,p_\mu \left(\sqrt{1- K q^2}-\frac{K p\cdot  q}{ \left(1+\sqrt{1-K p^2}\right)}  \right) + q_\mu  \,,
\label{DCLSnyder-1_curved}
\ee
where $p\cdot  q=p_\mu q_\nu a^{\mu\nu}(x)$, $p^2=p_\mu p_\nu a^{\mu\nu}(x)$ and $q^2=q_\mu q_\nu a^{\mu\nu}(x)$.

This construction would lead to a non-commutative structure (spacetime) attached to each point of the classical spacetime. As in the flat spacetime case, different local non-commutative algebras on curved spacetimes  can be constructed by the redefinition of the translation generators, as introduced in equation \eqref{eq:transredef2}. The redefinition of the translation generators can depend on the space-time region, and thus, it is possible to construct different local non-commutative structures in different regions on spacetime.

For the $\kappa$-Poincar\'e algebra in the classical basis on curved spacetimes the generators of translation \eqref{eq:transkappacurved} satisfy
\be
\begin{split}
	[ \tilde{\mathcal{T}}^\alpha,  \tilde{\mathcal{T}}^\beta] \,=\,&\frac{Z_\gamma}{\Lambda}\left( \tilde{\mathcal{T}}^\alpha a^{\beta\gamma} - \tilde{\mathcal{T}}^\beta a^{\alpha\gamma} \right) \,,\\ [ \tilde{\mathcal{T}}^\alpha, \mathcal{J}^{\beta\gamma}]\,=\,& a^{\alpha\beta}  \tilde{\mathcal{T}}^\gamma - a^{\alpha\gamma}  \tilde{\mathcal{T}}^\beta+\frac{Z_\delta}{\Lambda}\left(\eta^{\delta\beta} \mathcal{J}^{\alpha\gamma}-a^{\delta\gamma} \mathcal{J}^{\alpha\beta}\right)\,.
\end{split}
\label{cov_algebra_kappa_curved}
\ee

Mathematically precise, one can say that the non-commuting translations on each momentum space at each point in spacetime define local non-commutativity. In the flat spacetime case, due to the existence of a global coordinate system, one can identify momentum spaces with spacetime itself globally, and thus the non-commutativity of the momentum spaces is inherited to the flat spacetime globally. On a curved spacetime this can only be done locally. Our approach, starting from the differential geometry of a curved spacetime and implementing a local non-commutative structure on its momentum spaces (technically cotangent spaces), is a complementary approach to the algebraic one, which is performed for maximally symmetric spacetimes in \cite{Ballesteros:2019hbw}.

We like to point out that, from a classical differential geometry of curved manifolds point of view, this is a very natural construction: to deform the local cotangent/tangent space structure to introduce a new geometric structure on curved spacetime, such as non-commutativity. And also, from a physical point of view, the localization of symmetry algebras is precisely what is done in gauge field theories in particle physics, and what happens in the transition from special to general relativity, when one passes from global to local Lorentz transformations as symmetries of the theory. 

Hence, in our opinion, the just outlined approach to non-commutative curved spacetimes, which starts from a curved spacetime and localizes a non-commutative structure to each point of this spacetime, has strong prospects to yield new insights in the description of quantum deformations of the Poincar\'e algebra and non-commutativity on curved spacetimes.

The full analysis of this new program will be investigated in a series of future articles. We start in this article by studying consistency conditions and consequences from deformed relativistic kinematics on curved spacetimes on the curved momentum space and spacetime, i.e., curved phase-space geometry.

\section{Consistent phase space geometry, purely from a momentum space metric}\label{sec:GeomT*M}

In this section we identify self-consistent momentum space geometries, which are based on a position dependent momentum space metric, encoding DRKs at each point of spacetime.

We start by summarizing the main geometrical framework to describe the geometry of curved momentum spaces on curved manifolds consistently, following~\cite{miron2001geometry,2012arXiv1203.4101M}. Mathematically speaking, we are looking at the geometry of the cotangent bundle of a manifold defined by a specific cotangent bundle metric. Physically speaking, we are looking at the geometry of the point particle phase space.

Afterwards we identify those cotangent bundle geometries, which are compatible with the following two principles:
\begin{enumerate}
	\item The dispersion relation of physical point particles is defined by the minimal geometric distance in momentum space, determined by the momentum space metric.
	\item Solutions of the Hamilton equations of motion, determined by the dispersion relation defining Hamilton function, are horizontal curves, i.e., they are adapted to the geometry such that they can be interpreted as force-free, purely geometrically determined, particle trajectories.
\end{enumerate}
These two principles lead to several compatibility conditions, as has been pointed out in~\cite{Relancio:2020zok,Relancio:2020rys}, which we recall in Sec. \ref{ssec:geomMet}. 

The first principle was originally suggested in the context of relative locality~\cite{AmelinoCamelia:2011bm}. However, since our starting point is a maximally symmetric momentum space metric and the deformed kinematics obtained from it, it actually follows that any function of the metric distance in momentum space must be a Casimir element of the deformed relativistic symmetry algebra given by the isometries of the metric. Thus any of the Casimir operators can be chosen as mass operator, i.e.,\ dispersion relation. Simplicity, and a smooth limit to special and general relativity, suggests to use the square of of the metric distance~\cite{Carmona:2019fwf}.

The second point states that the point particle trajectories shall be compatible with the dispersion relation and the geometry which is derived from the momentum space metric.

We will investigate explicitly which kind of cotangent bundle geometries are compatible with these consistency conditions on a perturbative level in Sec. \ref{ssec:nT*Mgeom}, where we find that only certain position dependent momentum space metrics lead to a self-consistent geometry of the cotangent bundle realizing the two criteria.

All of this geometric analysis is done for general momentum space metrics and prepares the study of the consequences for the implementation of DRKs on curved spacetime based on maximally symmetric momentum spaces in Sec. \ref{sec:geomex}.

\subsection{The geometry of the cotangent bundle}
To discuss the geometry of a curved spacetime with curved momentum spaces we need the following notions, see for example \cite{miron2001geometry,2012arXiv1203.4101M}. 

\subsubsection{General notions}
Let $M$ be a smooth $n$-dimensional manifold (usually in physics is chosen as $n=4$). At each point $p\in M$ one can consider the cotangent spaces $T^*_pM$, whose union over the whole manifold form the so-called cotangent bundle $T^*M = \bigcup_{p\in M}T^*_pM$. In the following we will consider $T^*M$ in manifold induced coordinates, which are constructed as follows. A local coordinate chart $(U,x^\mu)$ on $M$ induces a local coordinate chart on $T^*U$ by identifying $u\in T^*U$ with the coordinates $(x,k)$ obtained from its coordinate expression $u=k_\mu dx^\mu\in T_x^*U$. The cotangent bundle is itself a $2n$-dimensional manifold and naturally carries the structure of a fibre bundle with local fibres $\mathbb{R}^n$, which are identified with the cotangent spaces $T^*_xM$. The bundle projection is given by $\pi:T^*M \to M; (x,k)\mapsto x$. 

The local coordinate bases of the tangent $T_{(x,k)}T^*M$ and cotangent spaces $T^*_{(x,k)}T^*M$ of $T^*M$ will be denoted by 
\begin{align}
	\lbrace \partial_\mu\,=\,\tfrac{\partial}{\partial x^\mu} ,\ \bar{\partial}^\mu\,=\,\tfrac{\partial}{\partial k_\mu}\rbrace\qquad \textrm{ and } \qquad \{dx^\mu, dk_\mu\}\,,
\end{align}
respectively.

Our aim is to set up a geometry of the cotangent bundle $T^*M$ such that at each point a clear split between the base manifold $M$ (position space) and the cotangent spaces $T^*_xM$ (momentum spaces) is ensured. This can be done in a mathematical precise way with the help of a so-called non-linear connection on $T^*M$.

\subsubsection{The non-linear connection}
The tangent and cotangent spaces of $T^*M$ can be split into vertical and horizontal subspaces, which are physically interpreted as tangent spaces to momentum and position space, respectively.

The vertical tangent spaces at a point $\mathcal{V}_{(x,k)}$ is canonically defined as $\ker d\pi_{(x,k)}$ and is nothing but the tangent space to the fibre $T^*_xM$. In a local coordinate basis $\mathcal{V}_{(x,k)} = \mathrm{span}\left\{\bar\partial^\mu\right\}$. The union of all vertical spaces $\mathcal{V} = \bigcup_{(x,k)\in T^*M}\mathcal{V}_{(x,k)}$ is called the vertical tangent bundle of $T^*M$.

The whole tangent space $T_{(x,k)}T^*M$ can then be split into its vertical part $\mathcal{V}_{(x,k)}$ and a complement $\mathcal{H}_{(x,k)}$, called the horizontal tangent space. The union of all horizontal spaces $\mathcal{H} = \bigcup_{(x,k)\in T^*M}\mathcal{H}_{(x,k)}$ is called the horizontal tangent bundle of $T^*M$. 

The freedom in defining the horizontal space is encoded in the choice of a connection on $T^*M$, defined by local connection coefficients $N_{\nu\mu}(x,k)$, which are needed to construct the local basis of $\mathcal{H}_{(x,k)} = \mathrm{span}\left\{\delta_\mu\right\}$ as 
\begin{equation}
	\delta_\mu\, = \,\partial_\mu + N_{\nu\mu}(x,k) \bar{\partial}^\nu\,.
	\label{eq:delta_derivative}
\end{equation}  
The main important property of these basis elements, which defines the transformation behaviour of the non-linear coefficients, is that, under manifold induced coordinate transformations, they transform tensorial, analogously to the $\partial_\mu$ basis of $T_xM$ under coordinate changes on the base manifold, i.e.,\
\begin{align}
	x\mapsto \tilde x(x) \Rightarrow \delta_a \mapsto \tilde \delta_\mu\, =\, \tilde \partial_\mu x^\nu \delta_\nu\,.
\end{align}
This transformation behaviour makes them basis for so-called distinguished or d-tensor on $T^*M$, which are tensors on $T^*M$ whose components behave under manifold induced coordinate transformations analogously to tensor components of tensors fields on the base manifold $M$.

In summary, with help of a connection one can split the tangent spaces $T_{(x,k)T^*M}$ of $T^*M$ into horizontal and vertical subspaces
\begin{align}
	T_{(x,k)}T^*M \,=\, \mathcal{V}_{(x,k)} \oplus \mathcal{H}_{(x,k)} \,=\, \mathrm{span}\left\{\bar\partial^\mu\right\} \oplus \mathrm{span}\left\{\delta_\mu\right\}\,,
\end{align}
where the vertical space can be identified with the tangent spaces to $T^*_xM$ (physically to the momentum spaces) and the horizontal space can be identified with the tangent spaces to $M$ (physically to position space). An analogous split for the cotangent spaces $T^*_{(x,k)T^*M}$ of $T^*M$ exists and is written as
\begin{align}
	T^*_{(x,k)}T^*M \,=\, \mathcal{V}^*_{(x,k)} \oplus \mathcal{H}^*_{(x,k)}\, =\, \mathrm{span}\left\{\delta k_\mu\right\} \oplus \mathrm{span}\left\{dx^\mu\right\}\,,
\end{align}
with
\begin{align}
	\delta k_\mu \,=\, d k_\mu - N_{\nu\mu}(x,k)\,dx^\nu\,. 
\end{align}

The choice of the connection coefficients defines the geometry of the cotangent bundle. In general, the $N_{\nu\mu}(x,k)$ can have a non-linear dependence on $k$ and are called non-linear connection coefficients. In the case of the existence of an affine connection on the base manifold $M$, defined by local connection coefficients $\Gamma^\sigma{}_{\mu\nu}(x)$, these define linear connections on $T^*M$ through the connections coefficients 
\begin{align}
	N_{\nu\mu}(x,k) \,=\, \Gamma^\rho{}_{\nu\mu}(x)k_\rho\,.	
	\label{eq:nonlinear_connection}
\end{align}
For pseudo-Riemannian manifolds, which are equipped with a spacetime metric $a = a_{\mu\nu}(x)dx^\mu \otimes dx^\nu$, $\Gamma^\rho{}_{\mu\nu}(x)$ can for example be chosen as the Christoffel symbols of the Levi-Civita connection of $a$.

\subsubsection{The non-linear curvature}
Connections immediately lead to the notion of curvature, which measure the integrability of the tangent spaces and can be related to the nonlinear connection coefficients, as we will see in the following. The non-linear curvature of the non-linear connection is defined as
\begin{equation}
	R_{\rho\mu\nu}(x,k)\bar\partial^\rho\,=\,\left[ \delta_\mu\,,\delta_\nu\right]\,=\,\left(\delta_\mu  N_{\rho\nu}(x,k)- \delta_\nu N_{\rho\mu}(x,k)\right)\bar\partial^\rho\,.
	\label{eq:dtensor}
\end{equation} 
Physically, this object represents the curvature of spacetime in phase space and is in general position and momentum dependent. In case the non-linear connection is linear in the momenta $k$, it is related to the Riemann curvature tensor ${R^{\sigma}}_{\rho\mu\nu}(x)$ of an affine connection on the base manifold $M$ 
\begin{equation}
	R_{\rho\mu\nu}(x,k)\,=\,k_\sigma {R^{\sigma}}_{\rho\mu\nu}(x)\,.
\end{equation} 
In virtue of Frobenius theorem, the non-linear curvature measures the integrability of spacetime, i.e., position space, as a subspace of the cotangent bundle. 

\subsubsection{The metric and its compatible affine connection}
After the discussion of the split of the tangent spaces of the cotangent bundle into position (horizontal) and momentum (vertical) space parts, a metric on the cotangent bundle, which defines a position and momentum space metric, can be defined as
\begin{equation}
	\mathcal{G}\,=\, g_{\mu\nu}(x,k) dx^\mu dx^\nu+g^{\mu\nu}(x,k) \delta k_\mu \delta k_\nu\,.
	\label{eq:line_element_ps} 
\end{equation}
It makes $T^*M$ a metric manifold and we can determine metric compatible affine connections on $T^*M$. The Levi-Civita connection of \eqref{eq:line_element_ps} has the drawback that, in general, it does not respect the horizontal-vertical split of the non-linear connection, i.e., its covariant derivative does not map vertical vectors to vertical ones, or horizontal vectors to horizontal ones. However, there exists a metric compatible connection which does so~\cite{miron2001geometry,2012arXiv1203.4101M}. It is defined by the covariant derivative operations
\begin{align}
	&\nabla_{\delta_\mu} \delta_\nu \,=\, H^\sigma{}_{\mu\nu}(x,k) \delta_\sigma,\quad \nabla_{\delta_\mu} \bar\partial^\nu\, =\, -H^\nu{}_{\mu\sigma}(x,k) \bar\partial^\sigma\,, \\
	&\nabla_{\bar\partial^\mu} \bar\partial^\nu\, =\,- C_\sigma{}^{\mu\nu}(x,k)\bar\partial^\sigma,\quad \nabla_{\bar\partial^\mu} \delta_\mu \,=\, C_\mu{}^{\sigma\nu}(x,k)\delta_\sigma\,,
\end{align}
where the affine connection coefficients are given by
\begin{align}
	{C_\rho}^{\mu\nu}(x,k)\,=\,-\frac{1}{2}g_{\rho\sigma}\left(\bar{\partial }^\mu g^{\sigma\nu}(x,k)+ \bar{\partial }^\nu g^{\sigma\mu}(x,k)-\bar{\partial }^\sigma g^{\mu \nu}(x,k)\right)\,,
	\label{eq:affine_connection_p}\\
	{H^\rho}_{\mu\nu}(x,k)\,=\,\frac{1}{2}g^{\rho\sigma}(x,k)\left(\delta_\mu g_{\sigma\nu}(x,k) +\delta_\mu g_{\sigma\mu}(x,k) -\delta_\sigma g_{\mu\nu}(x,k) \right)
	\label{eq:affine_connection_st}\,.
\end{align}
In our later discussion, two sets of curves will be of physical importance:
\begin{itemize}
	\item Vertical autoparallels of this metric compatible affine connection, i.e.\ curves $\gamma(\tau) = (x_0,k(\tau))$ satisfying $\nabla_{\dot\gamma}\dot \gamma = 0$, are solutions of the equations
	\begin{equation}
		\ddot k_\mu  - C_{\mu}{}^{\nu\sigma}(x,k) \dot{k}_\nu \dot{k}_\sigma\,=\,0\,.
		\label{eq:vertical}
	\end{equation} 
	They will define the distance in momentum space from which the dispersion relation is obtained.
	\item Horizontal autoparallels are curves $\gamma(\tau) = (x(\tau),k(\tau))$ characterized by the horizontallity condition
	\begin{equation}
		\delta \dot{k}_\lambda\,=\,\dot{k}_\lambda-N_{\sigma\lambda} (x,k)\dot{x}^\sigma\,=\,0\,,
		\label{eq:horizontal_momenta}
	\end{equation} 
	and the autoparallel equation
	\begin{equation}
		\ddot{x}^\mu+{H^\mu}_{\nu\sigma}(x,k) \dot{x}^\nu\dot{x}^\sigma\,=\,0\,.
		\label{eq:horizontal_geodesics_curve_definition}
	\end{equation} 
	They define force-free particle motion along spacetime and will be satisfied by solutions of the Hamilton equations of motion defined by the dispersion relation.
\end{itemize}

\subsection{Mathematical realization of the principles}\label{ssec:geomMet}
We introduced all the notions needed to cast the two principles, which are listed at the beginning of Sec. \ref{sec:GeomT*M} in a precise mathematical statement. They imply non-trivial constraints on the cotangent bundle geometry in case the distance in momentum space, which is interpreted as dispersion relation of the point particles, is not a homogeneous function in the momenta.

\subsubsection{The Hamilton function and the dispersion relation}
Our first principle states that the dispersion relation is given by the geodesic distance in momentum space between the origin and a given momentum $k$, defined by the momentum space metric.

Consider a momentum space curve $k(\tau)$ with $k(0)=0$ and $k(\tau_1)=k$. For these curves, the geometric length measure defined by the metric on momentum space is
\begin{align}\label{eq:vertgeomdist}
	D(x,k) \,=\, \int_0^{\tau_1} d\tau \sqrt{g^{\mu\nu}(x,k(\tau))\dot k_\mu(\tau)\dot k_{\nu}(\tau)}\,.
\end{align}
Then, to extremize this length measure, it turns out that $k(\tau)$ has to satisfy
\begin{align}\label{vert:geod}
	\ddot k_\mu + \frac{1}{2}g_{\mu\sigma}(x,k)\left(\bar\partial^\rho g^{\sigma \lambda}(x,k) + \bar\partial^\lambda g^{\sigma \rho}(x,k) - \bar\partial^\sigma g^{\rho \lambda}(x,k) \right) k_\rho k_\lambda \,=\, 0\,,
\end{align}
which means that $k(\tau)$ is a vertical autoparallel, see \eqref{eq:vertical}.

The dispersion relation defining Hamilton function $\mathcal{C}(x,k)$, or mass Casimir operator of a symmetry algebra, can be identified with the square of the geometric distance $\mathcal{C}(x,k)=D(x,k)^2$, as has been shown in \cite{Relancio:2020rys}. In fact, as commented in Sec. \ref{sec:GeomT*M}, any function of $f(D(x,k))$ is a Casimir of the symmetry algebra obtained from the momentum space isometries. The identification with the square of the distance was considered as the simplest way to have a smooth limit to special and general relativity.

In \cite{Bhattacharya2012RelationshipBG,Relancio:2020zok}, it has been demonstrated that, to determine the expression for $\mathcal{C}(x,k)$, one can solve the following differential equation instead of solving the integral \eqref{eq:vertgeomdist} explicitly,
\begin{equation}
	\mathcal{C}(x,k)\,=\,\frac{1}{4}\bar{\partial}^\mu \mathcal{C}(x,k) g_{\mu\nu} (x,k)\bar{\partial}^\nu \mathcal{C}(x,k) \,.
	\label{eq:casimir_metric}
\end{equation} 
This equation is the first necessary condition which relates the dispersion relation and the momentum space metric. It is valid on flat, as well as, on curved spacetime.

The Hamilton function then defines the dispersion relation, the position and momentum of a physical particle have to satisfy
\begin{align}
	\mathcal{C}(x,k) \,=\, m^2\,.
\end{align}

\subsubsection{Particle motion}
The second principle states that the Hamilton equations of motion determined by the dispersion relation defining Hamilton function,
\begin{align}\label{eq:HamEom}
	\dot k_\mu + \partial_\mu \mathcal C (x,k)\, =\, 0,\quad \dot x^\mu \,=\, \bar\partial^\mu \mathcal{C} (x,k)\,,
\end{align}
shall be horizontal curves, so that they are adapted to the geometry and can be interpreted as force-free, purely geometrically determined, particle trajectories. The first Hamilton equation of motion can be rewritten in terms of the non-linear connection to take the form
\begin{align}
	\dot k_\mu - N_{\nu\mu}(x,k)\bar\partial^\nu \mathcal{C} + \delta_\mu \mathcal{C}(x,k)\, =\, 0\,.
\end{align}
Comparing this equation with the horizontallity condition \eqref{eq:horizontal_momenta} it is clear that solutions of the Hamilton equations of motion are horizontal curves if and only if the Hamiltonian satisfies
\begin{equation}
	\delta_\mu \mathcal{C}(x,k)\,=\,0\,.
	\label{eq:casimir_delta}
\end{equation} 
This condition connects the non-linear connection with the Hamiltonian and, in the virtue of \eqref{eq:casimir_metric}, with the momentum space metric.

For Hamilton functions which are positively $r$-homogeneous in $k$, i.e.\ $H(x,\lambda k) = \lambda^r H(x,k)$, with $\lambda >0$, this condition can always be satisfied for a specific choice of a canonical non-linear connection, which always exists and is uniquely constructed from the Hamiltonian alone, as it is known from the framework of Hamilton geometries (see \cite{miron2001geometry,Barcaroli:2015xda}). Hence, there always exists a self consistent geometry of to cotangent bundle for position dependent momentum space metrics, which leads via \eqref{eq:vertgeomdist} and \eqref{eq:casimir_metric} to homogeneous Hamiltonian functions, that automatically encode force-free, pure geometric particle motion. In general, if the resulting Hamiltonian is not homogeneous, this is not the case. Then, \eqref{eq:casimir_delta} is a non-trivial constraint which can be used to determine parts of the non-linear connection from the momentum space metric. In the framework of generalized Hamilton spaces, whose geometry is based on a position dependent momentum space metric (without any use of the Hamiltonian), a general solution for a non-linear connection such that \eqref{eq:casimir_delta} is satisfied is not known.

Evaluating \eqref{eq:casimir_delta} on the first condition \eqref{eq:casimir_metric} implies immediately another consistency constraint, which is
\begin{equation}
	{H^\rho}_{\mu\nu}(x,k)\,=\,\bar{\partial}^\rho N_{\mu\nu}(x,k)\,.
	\label{eq:affine_connection_n}
\end{equation} 
It connects the affine connection coefficients ${H^\rho}_{\mu\nu}(x,k)$ on $T^*M$ with the more fundamental non-linear connection coefficients $N_{\mu\nu}(x,k)$. Using this additional constraint in the second Hamilton equation of motion, $\dot x^\mu = \bar\partial^\mu \mathcal{C}$, implies the horizontal geodesic Eq.~\eqref{eq:horizontal_geodesics_curve_definition}.

The geometric construction presented so far  makes the solutions of the Hamilton equation of motion horizontal autoparallels of the metric compatible affine connection, as has also been shown in \cite{Relancio:2020rys}. 

In the following, we use a perturbative ansatz for the momentum space metric to determine the consequences from the compatibility constraints on the metric. If the momentum space metric components $g^{\mu\nu}(x,k)$ are independent of $k$, both conditions \eqref{eq:casimir_delta} and $\eqref{eq:affine_connection_n}$ are satisfied for non-linear connection coefficients \eqref{eq:nonlinear_connection}, which are generated by the Christoffel symbols of the Levi-Civita connection that are derived from the metric components $g^{\mu\nu}(x,k)=a^{\mu\nu}(x)$. We will show that this is not the only possible solution, but that there exists $k$-dependent momentum space metrics which satisfy \eqref{eq:casimir_delta} and $\eqref{eq:affine_connection_n}$. However, among all possible momentum space metrics which one may consider, in particular among those which are employed to encode DRKs, it turns out that only specific classes satisfy the constraints.

\subsection{\texorpdfstring{$n$}{n}-th order polynomial perturbative cotangent bundle geometry from a metric}\label{ssec:nT*Mgeom}
Deformations of the kinematics of high-energetic particles are expected to become relevant for particles of an energy near a high energy scale $\Lambda$, for example the Planck scale in the context of quantum gravity. To study such modifications, we make a first order perturbative expansion for the geometric objects involved that is polynomial in the momenta. We evaluate the compatibility conditions which relate the metric and the Hamiltonian~\eqref{eq:casimir_metric} as well as the non-linear- and the affine connection,~\eqref{eq:casimir_delta} and~\eqref{eq:affine_connection_n}.

The parameter $\epsilon$ below is a perturbation parameter which labels the first order non-vanishing deformation of the quantity under consideration. In the context of DRKs the deformation parameter is  given by $\epsilon = \frac{1}{\Lambda^q}$, where $q$ denotes the order of the polynomial deformation.

For the momentum space metric we use
\begin{equation}
	g^{\mu\nu}(x,k)\,=\,a^{\mu\nu}(x)+ \epsilon b^{\mu\nu (\rho_1 \cdots \rho_n)}(x)  k_{\rho_1} \cdots k_{\rho_n}\,,
	\label{eq:metric_1_up}
\end{equation}
whose inverse (indices are raised and lowered with the zeroth order metric components $a^{\mu\nu}$ and $a_{\mu\nu}$, respectively) is given by
\begin{equation}
	g_{\mu\nu}(x,k)\,=\,a_{\mu\nu}(x)- \epsilon {b_{\mu\nu}}^{(\rho_1 \cdots \rho_n)}(x)  k_{\rho_1} \cdots k_{\rho_n}\,,\label{eq:metric_1_down}
\end{equation}
and the Hamiltonian will be expressed as
\begin{equation}
	\mathcal{C}(x,k)\,=\, k_\mu k_\nu\left( A^{(\mu\nu)}(x)+ \epsilon B^{(\mu\nu\rho_1 \cdots \rho_n)}(x)  k_{\rho_1} \cdots k_{\rho_n}\right)\,.
	\label{eq:hamiltonian_1}
\end{equation}
For the nonlinear connection coefficients we use an ansatz of the form
\begin{equation}
	N_{\mu\nu}(x,k)\,=\,  k_\sigma \left({\Gamma^{\sigma}}_{\mu\nu}(x)+\epsilon {X_{\mu\nu}}^{(\sigma \rho_1 \cdots \rho_n)}(x)  k_{\rho_1} \cdots k_{\rho_n}\right) \,.
	\label{eq:nonlinear_1}
\end{equation}

Here $a^{\mu\nu}(x)$ is a Lorentzian spacetime metric, as employed in general relativity, and $\Gamma^{\sigma}{}_{\mu\nu}(x)$ are the affine connection coefficients of its Levi-Civita connection. The $A^{(\mu\nu)}(x)$ components define the zeroth order Hamilton function, a priori independently of the metric. The space-time tensor fields $b^{\mu\nu (\rho_1 \cdots \rho_n)}(x)$, $B^{(\mu\nu\rho_1 \cdots \rho_n)}(x)$ and $X_{\mu\nu}{}^{(\sigma \rho_1 \cdots \rho_n)}(x)$ parametrize the perturbations from Lorentzian spacetime geometry.

In the following we determine relations between these tensor fields from the conditions~\eqref{eq:casimir_metric},~\eqref{eq:casimir_delta} and~\eqref{eq:affine_connection_n}. For the sake of readability we omit to display the $x$ dependence of the spacetime tensors explicitly in the following calculations, wherever the dependence of the objects involved should be clear from the context.

\subsubsection{Compatibility between Hamiltonian and metric}
Let us first study the condition~\eqref{eq:casimir_metric}. Expanding this expression to first order in $\epsilon$ yields
\begin{align}
	&a_{\mu\nu} k_\sigma k_\alpha A^{\mu\sigma}A^{\nu\alpha}\\
	&+ \epsilon \left( a_{\mu\nu}A^{\mu\sigma}(n+2)B^{(\nu\alpha\beta_1...\beta_n)}k_\alpha k_\sigma  k_{\beta_1} ... k_{\beta_n}
	- A^{\mu\sigma}A^{\nu\alpha}{b_{\mu\nu}}^{(\lambda_1 ... \lambda_n)}  k_{\lambda_1} ... k_{\lambda_n}k_\sigma k_\alpha\right)\nonumber\\
	&=\,k_\mu k_\nu\left( A^{(\mu\nu)}+ \epsilon B^{(\mu\nu\rho_1 \cdots \rho_n)}   k_{\rho_1} \cdots k_{\rho_n}\right)\,,
\end{align}
which determines the coefficients in the Hamiltonian from the metric coefficients order by order. Equating the powers in the polynomial in $k$ yields at
zeroth order
\begin{align}\label{eq:A=a}
	A^{(\mu \nu)}\,=\,a^{\mu \nu}\,,
\end{align}
while the first order implies
\begin{align}
	B^{(\mu \nu \rho_1 \cdots \rho_n)}  \,=\,\frac{1}{n+1}b^{(\mu \nu \rho_1 \cdots \rho_n)}\,.\label{eq:metric_casimir:conditions}
\end{align}

\subsubsection{Compatibility between the non-linear connection and the Hamiltonian}
Having obtained the Hamiltonian from the metric, we expand \eqref{eq:casimir_delta}, which, in turn, intertwines the non-linear connection coefficients and the components of the metric. Expanding \eqref{eq:casimir_delta} to first order in $\epsilon$ and using \eqref{eq:nonlinear_1} yields
\begin{align}
	{X_{\mu\lambda}}^{(\sigma \rho_1 ... \rho_n)} A^{(\lambda\nu)}k_\nu k_\sigma  k_{\rho_1} ... k_{\rho_n}\, =\, - \frac{1}{2} \mathring{\nabla}_\mu B^{(\sigma\nu\rho_1 \cdots \rho_n)}  k_\nu k_\sigma k_{\rho_1} ... k_{\rho_n}\,,
\end{align}
where $\mathring{\nabla}$ denotes the covariant derivative defined by the Levi-Civita connection of the metric $a$.

In order for this equation to be satisfied, using \eqref{eq:A=a} and $\eqref{eq:metric_casimir:conditions}$, we find that 
\begin{align}\label{eq:condition_x}
	{X_{\mu}}^{(\nu\sigma \rho_1 ... \rho_n)}(x) \,=\, - \frac{1}{2(n+1)}\mathring{\nabla}_\mu b^{(\sigma\nu\rho_1 \cdots \rho_n)}(x)\,.
\end{align}

\subsubsection{Compatibility between nonlinear connection and the affine connection}\label{sssec:affNlPert}
To investigate the condition \eqref{eq:affine_connection_n}, which imposes a relation between the non-linear and affine connection coefficients on the tangent bundle, we first note that
\begin{align}
	\delta_\sigma g_{\mu\nu}(x,k) \,=\, \partial_\sigma a_{\mu\nu} -\epsilon  (\mathring{\nabla}_\sigma b_{\mu\nu}{}^{(\rho_1...\rho_n)} + &\Gamma^\lambda{}_{\mu\sigma}b_{\lambda\nu}{}^{(\rho_1...\rho_n)} +\\ 
	&\Gamma^\lambda{}_{\nu\sigma}b_{\lambda\mu}{}^{(\rho_1...\rho_n)})k_{\rho_1}...k_{\rho_n}\,,
\end{align}
allowing us to express the affine connection coefficients \eqref{eq:affine_connection_st}  as
\begin{align}
	H^\sigma{}_{\mu\nu}(x,k) \,=\, \Gamma^\sigma{}_{\mu\nu} - \frac{1}{2}a^{\sigma\lambda}(\mathring{\nabla}_\mu b_{\lambda\nu}{}^{(\rho_1...\rho_n)} &+ \mathring{\nabla}_\nu b_{\lambda\mu}{}^{(\rho_1...\rho_n)}\\& - \mathring{\nabla}_\lambda b_{\mu\nu}{}^{(\rho_1...\rho_n)})k_{\rho_1}...k_{\rho_n}\,.
\end{align}
The $k$-derivative of the non-linear connection coefficients \eqref{eq:nonlinear_1} yields
\begin{equation}
	\bar\partial^\sigma N_{\mu\nu} (x,k)\,=\, \Gamma^{\sigma}{}_{\mu\nu}(x)+ (n+1)\epsilon X_{\mu\nu}{}^{(\sigma \rho_1 \cdots \rho_n)}(x)  k_{\rho_1} \cdots k_{\rho_n}\,.
\end{equation}
Now, by imposing condition~\eqref{eq:affine_connection_n} one gets 
\begin{align}
	{X_{\mu\nu}}^{(\sigma \rho_1 \cdots \rho_n)} k_{\rho_1}...k_{\rho_n}  \,=\,- \frac{1}{2(n+1)}a^{\sigma\lambda}(\mathring{\nabla}_\mu b_{\lambda\nu}{}^{(\rho_1...\rho_n)} +& \mathring{\nabla}_\nu b_{\lambda\mu}{}^{(\rho_1...\rho_n)} \\-& \mathring{\nabla}_\lambda b_{\mu\nu}{}^{(\rho_1...\rho_n)})k_{\rho_1}...k_{\rho_n}\,.
	\label{eq:x_11}
\end{align}
By contraction with $k_\sigma$ and $k_\nu$, one finds that this equation actually implies the compatibility~\eqref{eq:condition_x} we found earlier. 

Eq.~\eqref{eq:x_11} imposes an important constraint on the metric perturbation tensor $b$. Only those tensors $b$ which satisfy this constraint lead to a cotangent bundle geometry, which satisfies the principles listed in the beginning of Sec. \ref{sec:GeomT*M}. Since the left hand side is symmetric in the exchange of $\rho_i$ and $\sigma$, the right hand side must also satisfy this symmetry condition, which is not guaranteed for an arbitrarily chosen $b$.

To classify in general for which $b_{\lambda\nu}{}^{(\rho_1...\rho_n)}$ the constraint \eqref{eq:x_11} does not lead to a contradiction, is beyond the scope of this article. Certainly, there exist consistent momentum dependent perturbations of metric spacetime geometry, as for example
\begin{align}\label{eq:bNonTriv}
	b^{\mu\nu\rho} = a^{\mu\rho} a^{\nu\sigma} \partial_{\sigma} \phi + a^{\nu\rho} a^{\mu\sigma} \partial_{\sigma} \phi\,,
\end{align}
for $n=1$ demonstrates. It leads to $X_{\mu\nu}{}^{\sigma\rho} = - \frac{1}{2} a^{\sigma\rho}\nabla_\mu \partial_\nu \phi$ and the index symmetries of both sides of the equations match. This is not always the case, as we will see next.

\section{Deformed relativistic kinematics on curved spacetime with consistent cotangent bundle geometry}\label{sec:geomex}
In this section we will study the consequences from the compatibility conditions~\eqref{eq:casimir_metric},~\eqref{eq:casimir_delta}, and~\eqref{eq:affine_connection_n}, for cotangent bundle metrics which encode DRKs, in the sense discussed in Sec. \ref{sec:DRKsFlatToCurved}. 

This means we consider a maximally symmetric momentum space metric on a flat spacetime with components $\zeta^{\mu\nu}(k)$, which is lifted to a cotangent bundle metric with the help of the tetrads $e^\mu{}_\alpha(x)$ of a Lorentzian spacetime metric $a$ ($a^{\mu\nu}(x) = \eta^{\alpha\beta}e^\mu{}_\alpha(x) e^\nu{}_\beta(x)$) by the mapping
\begin{align}
	\zeta^{\mu\nu}(k) \to g^{\mu\nu}(x,k) = \zeta^{\alpha\beta}(\bar k(k)) e^\mu{}_\alpha(x) e^\nu{}_\beta(x)\,
\end{align}
where $\bar{k}_\alpha= e^{\mu}{}_\alpha k_\mu$, as it was already mentioned in~\eqref{eq:definition_metric_cotangent}. In general, the cotangent bundle metric depends on the tetrad one chooses.

Different momentum space bases of DRKs are encoded in different momentum space metrics, which thus lead to different cotangent bundle metrics. For the  majority of the models studied in the literature, the momentum space metric is constructed from the Kronecker delta $\delta^\mu_\nu$, the Minkowski metric $\eta_{\mu\nu}$, and a vector field with constant components $n^\nu$, where often $n$ is chosen as $n^\mu=(1,0,0,0)$~\cite{Carmona:2016obd}, in order to obtain an isotropic (rotational invariant) momentum space metric and thus  isotropically deformed kinematics. In the following, we study the consequences for such models.

We will see that only certain DRKs lead to cotangent bundle metrics, which define a self-consistent cotangent bundle geometry, i.e., which satisfy the constraints \eqref{eq:casimir_delta} and~\eqref{eq:affine_connection_n}.

\subsection{Constraints on the momentum coordinates for deformed relativistic kinematics on curved spacetime}\label{ssec:constraintDRKS}
The position dependent momentum space metric for a DRKs model under consideration can be expanded into the polynomial form \eqref{eq:metric_1_up}, which yields a specific perturbation tensors $b^{\mu\nu(\rho_1...\rho_n)}$. Using the construction~\eqref{eq:definition_metric_cotangent}, one finds the following relation 
\begin{equation}
	{b_{\mu\nu}}^{(\rho_1 \cdots \rho_n)}(x)\,=\,e^\alpha_\mu(x)e^\beta_\nu(x)e^{\rho_1}_{\gamma_1}(x) \cdots e^{\rho_n}_{\gamma_n}(x) {\bar{b}_{\alpha\beta}}{}^{(\gamma_1 \cdots \gamma_n)} \,,
	\label{eq:b_bb}
\end{equation}
where ${\bar{b}_{\alpha\beta}}{}^{(\gamma_1 \cdots \gamma_n)}$ is constructed from the Kronecker delta $\delta^\mu_\nu$, the Minkowski metric $\eta_{\mu\nu}$, and a vector field with constant components $n^\mu$, which is often chosen as $n^\mu=(1,0,0,0)$. Eq.~\eqref{eq:b_bb} entails that ${b_{\mu\nu}}^{(\rho_1 \cdots \rho_n)}(x)$ is constructed from the space-time metric $a$, the Kronecker delta $\delta^\mu_\nu$ and a vector field $Z = Z^\mu \partial_\mu= e^\mu{}_\sigma(x) n^\sigma \partial_\mu$, for the models we are interested in, in this section.

For the most prominent models in the literature we list the perturbation tensors and evaluate the compatibility condition \eqref{eq:x_11}.
\begin{itemize}
	\item The $\kappa$-Poincar\'e algebra in the bicrossproduct basis~\cite{Gubitosi:2013rna,Carmona:2019fwf} on curved spacetime 
	\begin{equation}
		b^{\mu\nu\rho} \,=\, 2 a^{\mu\nu}Z^\rho + 2 Z^\nu Z^\rho Z^\mu\,.
	\end{equation}
	Evaluating \eqref{eq:x_11} for this case yields
	\begin{align}
		X_{\mu\nu}{}^{(\sigma\rho)}\,=&\,\frac{1}{2}
		\left(\delta^\rho_\nu \mathring{\nabla}_\mu Z^\sigma+\delta^\rho_\mu \mathring{\nabla}_\nu Z^\sigma-a_{\mu \nu} \mathring{\nabla}^\rho Z^\sigma\right. \\
		&\left.+ \mathring{\nabla}_\mu (Z_\nu Z^\rho Z^\sigma)+ \mathring{\nabla}_\nu (Z_\mu Z^\rho Z^\sigma)- \mathring{\nabla}^\rho (Z_\nu Z_\mu Z^\sigma)\right)\,.	
	\end{align}
	The way the perturbation is constructed the left hand side must be symmetric in its upper indices, but the right hand side of the equation is clearly not symmetric in $\sigma$ and $\rho$. 
	Thus, the $\kappa$-Poincar\'e algebra DRKs in the bicrossproduct basis cannot be consistently lifted to curved spacetime, with the procedure based on the momentum space metric, which we discussed so far. 
	\item The Snyder algebra in the Maggiore realization~\cite{Battisti:2010sr} and the $\kappa$-Poincar\'e algebra in the classical basis~\cite{Borowiec2010} are based on a momentum space metric which we presented in \eqref{eq:gmaxcurved}~\cite{Carmona:2019fwf}. It leads to
	\begin{equation}
		b^{\mu\nu(\rho_1\rho_2)} = \frac{1}{2}(a^{\mu\rho_1}a^{\nu\rho_2} + a^{\mu\rho_2}a^{\nu\rho_1})\,.
	\end{equation}
	This deformation tensor  immediately satisfies \eqref{eq:x_11} since it is covariantly constant, i.e.\ $\mathring{\nabla}_\sigma b^{\mu\nu(\rho_1\rho_2)} = 0$ holds, and hence, these DRKs can be lifted to curved spacetime with the algorithm we presented, leading to a vanishing $X_{\mu\nu}{}^{\sigma\rho_1 \rho_2}$.
	\item In~\cite{Relancio:2020rys} we found a particular momentum basis for the DRKs, which leads to the following tensor field
	\begin{equation}
		b^{\mu\nu(\rho_1\rho_2)} = -\frac{1}{2}a^{\mu\nu}a^{\rho_1\rho_2} \,.
	\end{equation}
	By the same argument used in the previous example, $\mathring{\nabla}_\sigma b^{\mu\nu(\rho_1\rho_2)} = 0$, implying that  \eqref{eq:x_11} is automatically satisfied, being zero. 
\end{itemize}

In general, a covariantly constant perturbation tensor for which $\nabla_\sigma b^{\mu\nu(\rho_1...\rho_n)}=0$ is viable. As demonstrated by the first example of $\kappa$-Poincaré, Eq.~\eqref{eq:x_11} is not satisfied by a generic vector field $Z_\mu$ on any curved spacetime, i.e., for any space-time tetrad. There may exist particular spacetimes with high symmetry, for which there exist tetrads such that~\eqref{eq:x_11} can be satisfied. This however would lead to the fact that the DRKs under consideration cannot be implemented on generically curved spacetimes for a generic tetrad. Note that the non-trivial example displayed in \eqref{eq:bNonTriv}, which leads to a consistent solution of~\eqref{eq:x_11},  does not appear as perturbation tensor in the class of DRKs models under consideration.

To ensure that~\eqref{eq:x_11} holds for any choice of spacetime tetrad, for the perturbations we consider in this section, one needs that the tensor field $b^{\mu\nu(\rho_1...\rho_n)}$ is covariantly constant. This means that it is constructed from the components of the space-time metric $a^{\mu\nu}$ or the Kronecker delta $\delta^\mu_\nu$ alone, and that the components of the distinguished vector field $Z^\mu$, which cannot be covariantly constant for every possible tetrad from which it may be constructed, cannot appear. Thus, in the context of DRKs, the most general $b^{\mu\nu(\rho_1...\rho_n)}$ satisfying ~\eqref{eq:x_11} on any spacetime for any tetrad has the following form
\begin{equation}\label{eq:b=adelta}
	{b}_{\mu\nu}^{\quad (\rho_1 \cdots \rho_n)}\,=\, b_1 a_{\mu\nu} a^{(\rho_1\rho_2}...a^{\rho_{n-1}\rho_{n})} + b_2 \delta^{(\rho_1}_\mu  \delta^{\rho_2}_\nu a^{\rho_3\rho_4}...a^{\rho_{n-1}\rho_{n})}\,,
\end{equation}
where $b_1$ and $b_2$ are constants. Implications are that the polynomial power counting index $n=2N$ must be even, and that, when we use \eqref{eq:b=adelta} in \eqref{eq:metric_1_up}, the hole perturbatively deformed momentum space metric can be written as
\begin{align}\label{eq:metric_Lorentz}
	g_{\mu\nu}(x,k) = a_{\mu\nu}(x)  f_1\left(\frac{k^2}{\Lambda^2}\right) + \frac{1}{\Lambda^2}  k_\mu k_\nu f_2\left(\frac{k^2}{\Lambda^2}\right)\,,
\end{align}
where $k^2 =  a^{\mu\nu}(x)k_\mu k_\nu$, and the perturbation functions are $f_1 = 1+\epsilon b_1 (k^2)^N$ and $f_2 = \epsilon b_2 (k^2)^{N-1}$ with $\epsilon = \frac{1}{\Lambda^{2N}}$.

For such a metric, the corresponding non-linear connection coefficients are actually linear in the momenta and defined solely by the Christoffel symbols of the metric $a$, see \eqref{eq:nonlinear_1}. Consequently, by Eq.~\eqref{eq:affine_connection_n}, the affine connection coefficients are identical to the Christoffel symbols of the Levi-Civita connection of the metric $a$, and the horizontal part of the cotangent bundle geometry reduces to the usual Lorentzian metric space-time geometry defined by the metric.

However, at each point on spacetime, the cotangent spaces/momentum spaces still posses a non-trivial geometry determined by the metric \eqref{eq:metric_Lorentz}, which simultaneously encodes DRKs on - and local Lorentz invariance of - the curved space-time geometry.

Our perturbative analysis of the geometric consistency conditions (which were derived from the principles on the cotangent bundle geometry) shows that DRKs models cannot be lifted to arbitrary curved spacetimes for any momentum basis.  Indeed, we find that this lift can be done only with a momentum space basis leading to a local momentum space metric which is local Lorentz invariant, i.e., its lift to the cotangent bundle does not dependent on the tetrad chosen.

\subsection{Local Lorentz invariance as guiding principle}\label{ssec:constraintDRKsNonPert}
Our findings from the perturbative analysis lead to the conjecture of another guiding principle for the construction of deformed relativistic kinematics on curved spacetimes. As already mentioned several times, for general momentum space metrics, the lifting procedure to curved spacetimes leads to different cotangent bundle metrics for different tetrads. Given two tetrads $e^\mu{}_\alpha(x)$ and $\hat e^\nu{}_\beta(x) = \Lambda^\nu{}_\sigma(x) e^\sigma{}_\beta(x)$ of a spacetime metric $a$, where $\Lambda^\nu{}_\sigma(x)$ is a local Lorentz transformation,
\begin{align}
	g_{e}^{\mu\nu}(x,k) \,= \, e^\mu{}_\alpha(x) e^\nu{}_\beta(x) \zeta^{\mu\nu}(\bar k(k)) \,\neq\, g_{\hat e}^{\mu\nu}(x,k) \,=\, \hat e^\mu{}_\alpha(x) \hat e^\nu{}_\beta(x) \zeta^{\mu\nu}(\hat k(k))\,.
\end{align}
If we now demand that $g_{e}^{\mu\nu}(x,k) = g_{\hat e}^{\mu\nu}(x,k)$, i.e., that for a given momentum space metric the lifting procedure to curved spacetime is independent of the tetrad, this yields that the cotangent bundle metric is a function of $k^2 = a^{\mu\nu}(x,k)k_\mu k_\nu$ and that the components must be given by
\begin{align}\label{eq:metric_Lorentz_nonpert}
	g^{\mu\nu}(x,k) = a^{\mu\nu}(x)  h_1\left(\frac{k^2}{\Lambda^2}\right) + \frac{1}{\Lambda^2}  k^\mu k^\nu h_2\left(\frac{k^2}{\Lambda^2}\right)\,,
\end{align}
where the functions $h_1$ and $h_1$ can, in principle, be general functions in $k^2$. A quick calculation shows that for these metrics, in general, the conditions~\eqref{eq:casimir_metric},~\eqref{eq:casimir_delta} and~\eqref{eq:affine_connection_n} are satisfied.

\begin{itemize}
	\item Eq.~\eqref{eq:casimir_metric} can be solved explicitly for any metric of the form of~\eqref{eq:metric_Lorentz_nonpert}. Since this metric is a function of $k^2$, the Hamiltonian will also be. Being the square of the metric distance in momentum space, it shares the same symmetries. The generators of the isometries for the metric~\eqref{eq:metric_Lorentz_nonpert} contain the generators of the usual undeformed linear Lorentz transformations, hence also the Hamiltonian must be invariant under these, and thus it can only depend on linear Lorentz invariant terms, i.e., on $k^2$. With this, evaluating \eqref{eq:casimir_metric} yields
	\begin{align}
		4 \mathcal{C}(x,k)\,=\,&\bar{\partial}^\mu \mathcal{C}(x,k) g_{\mu\nu}(x,k) \bar{\partial}^\nu \mathcal{C}(x,k) \,=\, 4\left(\frac{\partial \mathcal C}{\partial k^2}\right)^2 k_\rho a^{\mu \rho}(x) k_\sigma a^{\nu \sigma}(x) g_{\mu \nu}(x,k) \nonumber\\
		\,=\,& 4\left(\frac{\partial \mathcal C}{\partial k^2}\right)^2 \left(k^2 h_1\left(\frac{k^2}{\Lambda^2}\right) +\frac{(k^2)^2 }{\Lambda^2} h_2\left(\frac{k^2}{\Lambda^2}\right) \right)\,.
	\end{align}
	The previous differential equation leads to the following expression for the dispersion relation defining Hamilton function 
	\begin{equation}
		\mathcal{C}(x,k)\,=\, \left(1+\frac{1}{2} \int^{k^2}_1 \frac{\Lambda}{\sqrt{\alpha(\Lambda^2 h_1(\alpha/\Lambda^2)+\alpha h_2(\alpha/\Lambda^2))}}\, d\alpha\right)^2\,.
		\label{eq:cas_cov}
	\end{equation}
	
	\item Evaluating Eq.~\eqref{eq:casimir_delta} for a Hamilton function $\mathcal{C}(x,k) = \mathcal{C}(k^2)$ implies that the non-linear connection coefficients must be of the form $N_{\mu\nu}(x,k) = \Gamma^\rho{}_{\mu\nu}k_\rho$. To see this we write
	\begin{align}\label{eq:Nofk1}
		\delta_{\mu}\mathcal{C}(k^2) 
		\, =\, \partial_{k^2}\mathcal{C}(k^2) \delta_{\mu} k^2 
		\,=\, \partial_{k^2}\mathcal{C}(k^2) (k_\rho k_\sigma\partial_{\mu} a^{\rho\sigma} + 2 k_{\nu}a^{\nu\sigma}N_{\nu\mu}) 
		\,=\, 0\,,
	\end{align}
	and hence, for a non-trivial Hamilton function the bracket must vanish. Taking another $k$-derivative of the bracket implies
	\begin{align}\label{eq:Nofk2}
		2 k_\sigma\partial_\mu a^{\lambda\sigma} + 2 a^{\lambda\sigma}N_{\mu\nu} + 2 k_{\nu}a^{\nu\sigma}\bar\partial^\lambda N_{\nu\mu}=0\,.
	\end{align}
	Contracting \eqref{eq:Nofk2} with $k^\mu$ and using the symmetry of $N_{\mu\nu}$ and \eqref{eq:Nofk1} implies that $\bar\partial^\lambda N_{\nu\mu}(x,k)$ must be independent of $k$ and can thus be written as $N_{\mu\nu}(x,k) = \Gamma^\rho{}_{\mu\nu}k_\rho$. Plugging this expression again into \eqref{eq:Nofk1} and \eqref{eq:Nofk2} yields that the coefficients $\Gamma^\rho{}_{\mu\nu}$ must be exactly the Christoffel symbols of the Levi-Civita connection of the metric $a$, as it is used in general relativity.
	\item With the findings from Eq.~\eqref{eq:casimir_delta}, Eq.~\eqref{eq:affine_connection_n} is satisfied, since for the metric~\eqref{eq:metric_Lorentz_nonpert}
	\begin{align}
		\delta_\sigma g_{\mu\nu}(x,k)\,=&\, h_1\left(\frac{k^2}{\Lambda^2}\right) \partial_\sigma a_{\mu\nu}(x,k)\\&+ \frac{1}{\Lambda^2}  h_2\left(\frac{k^2}{\Lambda^2}\right) k_\lambda \left({\Gamma^\lambda}_{\mu \sigma}(x) k_\nu +{\Gamma^\lambda}_{\nu \sigma}(x) k_\mu\right)\,,
		\label{eq:delta_metric}
	\end{align}
	holds. Then, using this in the definition of the spacetime affine connection~\eqref{eq:affine_connection_st} one can easily find that  
	\be
	{H^\lambda}_{\nu \sigma}(x,k)\,=\,{\Gamma^\lambda}_{\nu \sigma}(x)\,.
	\ee
	With this we show the compatibility conditions of Eqs.~\eqref{eq:affine_connection_n} and~\eqref{eq:affine_connection_st}.
\end{itemize}

Thus, demanding  the lifting procedure of DRKs encoding momentum space metrics $\zeta^{\mu\nu}(k)$ to curved spacetime to be independent of the choice of spacetime tetrad,   self-consistent cotangent bundle geometries (in the sense that they satisfy the principles listed in the beginning of Sec.~\ref{sec:GeomT*M}) must be linearly local Lorentz invariant.   The resulting space-time geometry (horizontal cotangent bundle geometry) is identical to the metric space-time geometry determined by the space-time metric $a$ whose tetrads $e$ were used for the lifting procedure. The momentum space geometry (vertical cotangent bundle geometry) is maximally symmetric and determined at each space-time point by \eqref{eq:metric_Lorentz_nonpert}. Therefore, we call the self-consistent cotangent bundle geometry defined by the metric $g^{\mu\nu}(x,k) =  e^\mu{}_\alpha(x) e^\nu{}_\beta(x) \zeta^{\mu\nu}(\bar k(k)) $ the $\zeta^{\mu\nu}$ induced quantum deformation of the spacetime $(M,a)$.

\section{Conclusions}
\label{sec:conclusions}
Starting from first principles we studied how deformed relativistic kinematics can be implemented consistently on curved spacetimes in terms of a locally maximally symmetric geometry of the cotangent bundle. Our starting point were three assumptions. The first we identified in Sec. \ref{ssec:DRKsCST}, while the other two we formulated as principles in the beginning of Sec.~\ref{sec:GeomT*M}. We cast  this assumptions into precise mathematical conditions using the framework a cotangent bundle geometry derived from a cotangent bundle metric which respects the horizontal/vertical (spacetime/momentum space) split, see Sec.s \ref{ssec:geomMet}. 

In summary our results are based on the following starting points:
\begin{enumerate}
	\item Given deformed relativistic kinematics encoded in a momentum space metric $\zeta(k) = \zeta^{\mu\nu}(k) dk_\mu \otimes dk_\nu$ on flat spacetime, its lift to a curved spacetime with metric $a(x) = \eta_{\mu\nu}e^\mu{}_\alpha(x) e^\nu{}_\beta(x)dx^\alpha \otimes dx^\beta$ is given by $g(x,k) = g^{\mu\nu}(x,k)\delta k_\mu \otimes \delta k_\nu$ where the components $g^{\mu\nu}(x,k)=e^\mu{}_\alpha(x) e^\nu{}_\beta(x) \zeta^{\alpha\beta}(\bar k)$ are generated from a tetrad $e^\mu{}_\alpha(x)$ of the metric $a$ and the flat spacetime momentum space metric $\zeta(\bar k)$ evaluated at $\bar k_\mu =e^\alpha{}_\mu(x)k_\alpha$, see \eqref{eq:definition_metric_cotangent}.
	\item The dispersion relation $\mathcal{C}(x,k)=m^2$ of physical point particles is defined by the minimal geometric distance in momentum space, which is defined by the momentum space metric through the relation $4 \mathcal{C}(x,k)\,=\bar{\partial}^\mu \mathcal{C}(x,k) g_{\mu\nu} (x,k)\bar{\partial}^\nu \mathcal{C}(x,k)$, see \eqref{eq:casimir_metric}.
	\item Solutions of the Hamilton equations of motion, determined by the dispersion relation defining Hamilton function $\mathcal{C}(x,k)$, are horizontal curves, i.e.,\ they are adapted to the geometry such that they can be interpreted as force-free, purely geometrically determined, particle trajectories. Mathematically, this lead to the conditions $\delta_\mu \mathcal{C}(x,k) = 0$ for the Hamilton function and ${H^\rho}_{\mu\nu}(x,k)\,=\,\bar{\partial}^\rho N_{\mu\nu}(x,k)$ for the non-linear and affine connection on $T^*M$, see Eqs.~\eqref{eq:casimir_delta} and~\eqref{eq:affine_connection_n}.
\end{enumerate}
Evaluating the constraints on a momentum dependent perturbation of metric space-time geometry encoded into a perturbation tensor $b$ in Sec. \ref{sssec:affNlPert}, we found that in particular point 3. leads to a strong constraint on the perturbation.

In Sec. \ref{sec:geomex} we considered flat momentum space metrics constructed from the Minkowski metric $\eta$, the identity matrix $\delta$ and a vector field $n$ with constant coefficients in a global Cartesian coordinate system, as they appear in the geometric description of the most studied DRKs in the literature. We found that they can be lifted to curved spacetimes in accordance with the points 1.-3. if there exists a momentum space basis such that the momentum space metric actually does not depend on the vector field $n$; in other words, if the DRKs are linear local Lorentz invariant. For such momentum space metrics, the horizontal geometry of the cotangent bundle, i.e., the geometry of spacetime, is identical to Lorentzian metric spacetime geometry on which general relativity is based, however the vertical geometry of the cotangent bundle, i.e., the momentum space geometry, is non-trivial. Consequently the phenomenology of these DRKs in the 1-particle sector is indistinguishable from general relativity and the non-trivial momentum space geometry on curved spacetime will only manifest itself in multi-particle processes. The precise phenomenology depends on the choice of translations generators, which define the modified law of momentum addition and the modified dispersion relation through the metric. This result should not be discouraging from the phenomenological point of view, but rather the opposite. The here presented geometrical approach to DRKs avoids the rather strong constraints, based on time delays of massless particles, and shows a path towards DRKs on curved spacetimes which are compatible with a high-energy deformation scale of TeV, as it has been discussed in the literature~\cite{Carmona:2017oit,Carmona:2018xwm,Carmona:2019oph,Relancio:2020mpa,Carmona:2020whi}.

In addition to the geometric considerations, we have addressed in Secs.~\ref{sec:noncom} and \ref{sec:noncomCST} that the non-commutativity of spacetime can be derived by the identification of translations generators in momentum space with non-commutative coordinates: globally on flat spacetime, locally on curved spacetimes. We have studied the particular cases of local Snyder and $\kappa$-Minkowski non-commutative structures. The further intensive investigation of this new conjecture how to describe non-commutative spacetimes from the differential geometric, curved spacetime perspective, as well as the construction of other models, like $\kappa$-Poincaré, is left for future works.

In the future it will be interesting to derive multi-particle scattering processes on curved spacetimes, to identify observables in which the DRKs on curved spacetimes we identified manifest themselves. One such process is for example the famous collisional Penrose process on Kerr spacetime, which certainly will be effected by DRKs.

We like to point out that surely there exits further self consistent curved momentum space geometries on curved spacetimes. In particular, another way to construct a consistent cotangent bundle geometry for DRKs on curved spacetime is not to insist on $\delta_\mu \mathcal{C}=0$, which would immediately circumvent our conclusion. This would imply that the Hamilton equations of motion will not lead to geodesic motion but include a force-like term. A possible interpretation would be that particles subject to DRKs of this type feel an effective force, a remnant of the underlying fundamental theory of quantum gravity, which prevent point particles from geodesic motion.

A huge class of cotangent bundle geometries which lead to, or are derived from, a homogeneous Hamilton function satisfy the $\delta_\mu \mathcal{C}=0$ part of principle 3. automatically, as it is know from the framework of Hamilton geometry \cite{Barcaroli:2015xda,miron2001geometry}. Then, it depends on whether the Hamilton function is derived from a momentum space metric or not, whether the connections have to satisfy ${H^\rho}_{\mu\nu}(x,k)\,=\,\bar{\partial}^\rho N_{\mu\nu}(x,k)$ or not. This class of geometries is in principle as preferable as the ones we identified in this article, however it is not clear if, or under which conditions, they encode self-consistent DRKs including a compatible deformed addition of momenta.

The same is true for the class of consistent cotangent bundle geometries that simply satisfy the constraints~\eqref{eq:casimir_delta} and~\eqref{eq:affine_connection_n}, of which we already presented an example generated by a non-constant scalar field on spacetime in \eqref{eq:bNonTriv}.

A next step in the analysis of the relation between self-consistent cotangent bundle geometries and DRKs is under which conditions, or how, the latter two geometries just mentioned can lead to self-consistent DRKs. Moreover, an open question to be investigated is if it is possible to construct DRKs from an arbitrary momentum metric on the cotangent spaces, and not only from maximally symmetric ones. The main difficulty to be overcome here lies in how to construct consistent deformed relativistic addition of momenta, when the momentum space metric has neither quasi-translations nor Lorentz-transformations as isometries. While in Ref.~\cite{Carmona:2019fwf} it was developed the simplest way to obtain a DRK from is a maximally symmetric momentum space, an idea to go beyond that is to start from a momentum space with more than four dimensions. In this case new generators should appear, and therefore, there would not be a so simple identification with the kinematical ingredients as explained here. Furthermore, in upcoming works, we will investigate possible dynamics for cotangent bundle geometries, and under which conditions they are identical to the Einstein equations or lead to modified theories of gravity.

With this article we systematically identified paths to lift DRKs to curved spacetimes and discussed the possibilities to describe them by a self-consistent cotangent bundle geometry.

\begin{acknowledgments}
CP was funded by the Deutsche Forschungsgemeinschaft (DFG, German Research Foundation) - Project Number 420243324. JJR acknowledges support from the INFN Iniziativa Specifica GeoSymQFT. We thank José Luis Cortés and Stefano Liberati for useful discussions. The authors would like to acknowledge networking support by the COST Action QGMM (CA18108).
\end{acknowledgments}

\bibliographystyle{utphys}
\bibliography{QuGraPhenoBib}

\end{document}